\documentclass[twocolumn,amsmath,aps]{revtex4}

\usepackage{amssymb,mathrsfs,bm,color,times}
\usepackage{graphicx,color}
\usepackage{bm}
\usepackage[hypertex]{hyperref}
\usepackage{amsmath}

\newcommand{\be}{\begin{equation}}
\newcommand{\ee}{\end{equation}}
\newcommand{\bea}{\begin{eqnarray}}
\newcommand{\eea}{\end{eqnarray}}
\newcommand{\bsube}{\begin{subequations}}
\newcommand{\esube}{\end{subequations}}

\newcommand{\Eq}[1]{Eq.\,(\ref{#1})}

\newcommand{\la}{\langle}
\newcommand{\ra}{\rangle}

\newcommand{\beq}{\begin{equation}}
\newcommand{\eeq}{\end{equation}}
\newcommand{\beqn}{\begin{eqnarray}}
\newcommand{\eeqn}{\end{eqnarray}}
\newcommand{\nl}{\nonumber \\}

\newcommand{\bsub}{\begin{subequations}}
\newcommand{\esub}{\end{subequations}}

\begin{document}
%\begin{CJK*}{GBK}{Song}

\title{Weak-value-amplification analysis beyond the AAV limit of weak measurements}

\author{Jianhua Ren}
\affiliation{Center for Joint Quantum Studies and Department of Physics,
School of Science, \\ Tianjin University, Tianjin 300072, China}

\author{Lupei Qin}
\affiliation{Center for Joint Quantum Studies and Department of Physics,
School of Science, \\ Tianjin University, Tianjin 300072, China}

\author{Wei Feng}
\affiliation{Center for Joint Quantum Studies and Department of Physics,
School of Science, \\ Tianjin University, Tianjin 300072, China}

\author{Xin-Qi Li}
\email{xinqi.li@tju.edu.cn}
\affiliation{Center for Joint Quantum Studies and Department of Physics,
School of Science, \\ Tianjin University, Tianjin 300072, China}

\date{\today}

%% \maketitle
\begin{abstract}
The weak-value (WV) measurement proposed by Aharonov, Albert and Vaidman (AAV)
has attracted a great deal of interest in connection with quantum metrology.
In this work, we extend the analysis beyond the AAV limit
and obtain a few main results.
{\it (i)}
We obtain non-perturbative result for the signal-to-noise ratio (SNR).
In contrast to the AAV's prediction, we find that the SNR asymptotically
gets worse when the AAV's WV $A_w$ becomes large,
i.e., in the case $g|A_w|^2>>1$, where $g$ is the measurement strength.
{\it (ii)}
With the increase of $g$ (but also small),
we find that the SNR is comparable to the result under the AAV limit,
while both can reach -- actually the former can slightly exceed -- 
the SNR of the standard measurement.
However, along a further increase of $g$, the WV technique will
become less efficient than the standard measurement,
despite that the postselection probability is increased.   
{\it (iii)}
We find that the Fisher information can characterize the estimate precision
qualitatively well as the SNR,
yet their difference will become more prominent with the increase of $g$.
{\it (iv)}
We carry out analytic expressions of the SNR in the presence of technical noises
and illustrate the particular advantage of the imaginary WV measurement.
The non-perturbative result of the SNR manifests a favorable range
of the noise strength and allows an optimal determination.
\end{abstract}

% \pacs{03.65.Yz,03.65.Sq,31.15.xv,31.15.xg}

\maketitle

\section{Introduction}

Quantum weak measurement with postselection was initially proposed
by Aharonov, Albert, and Vaidman (AAV) in their seminal work \cite{AAV88,AV90}.
The marked feature of this type of measurement is
that the resultant quantum weak values (WV)
can exceed the range of eigenvalues of the observable.
Despite the long time debate with theoretical curiosity,
the concept of WV has been found useful in quantum metrology, e.g.,
in developing novel schemes of quantum state tomography \cite{Lun11,Boy13,Boy14}
and for weak signal amplifications in parameter estimation
\cite{Kwi08,How09a,How09b,How10a,How10b,How13,Sim10,Hog11,Tur11,Ste11,Guo13,
How16,How17,Jor16,Jor17,Lun17,Du16,Ma16,Gho18},
while the latter application has been termed as
weak-value amplification (WVA) technique in literature.

As demonstrated in the early representative experiments, e.g.,
the first observation of the spin Hall effect of light \cite{Kwi08}
and measuring the small transverse deflections of an optical beam
with extremely high resolution \cite{How09a,How09b,How10a,How10b},
the WVA technique can lead to an amplification phenomenon,
just like a small image magnified by a microscope,
The amplification effect is of great interest from the experimental perspective,
since it gives access to an experimental sensitivity beyond the detector's resolution,
making thus possible to measure very small physical effects.
For the optical beam-deflection measurement, this technique also allows
to use high power lasers with low power detectors
while maintaining the optimal signal-to-noise ratio (SNR),
and holds the ability to obtain the ultimate limit
in deflection measurement with a large beam radius \cite{How09a,How09b,How10a,How10b}.
In regard to this similar advantage, it was also pointed out that
the WVA technique can outperform conventional measurement
in the presence of detector saturation \cite{Lun17}.
Finally and importantly, the WVA technique can have remarkable advantages
of reducing technical noise in some circumstances
\cite{Kwi08,How09a,How09b,Ste11,Sim10,How16,How17,Nish12,Ked12,Jor14}.

On the other hand, theoretical understanding highlights that the WVA technique
can put all of the information about the detected parameter
into a small portion of the events (after postselection) and claims that
this fact alone gives technical advantages \cite{Nish12,Ked12,Jor14}.
% {Ked12,Nish12}  % {Jor14,Kne14}  % {Tana13,FC14}
However, there existed controversial debates about
the WVA advantage against technical noises,
with diverse opinions and some negative comments \cite{Tana13,FC14,Kne14}.
It seems now clear that at least the scheme based on the imaginary WV measurement
has sound potentials to outperform standard measurement
in the presence of technical noise \cite{Ste11,Sim10,Ked12,Jor14},
e.g., by several orders of magnitude \cite{Sim10}.

%  {Jor08,Lor08,Wu11,Tana11,Fuj12,Susa12,Kof12} % {Qin15,Qin16,Qin17}
%   {Kor99,Kor11,BR-1,BR-2}
We notice that so far the large number of investigations
on the WVA merits analysis have been largely restricted
in the treatment under the AAV limit.
Despite that the generalized results of WV beyond the AAV limit
have been developed in different contexts and with different forms
\cite{Jor08,Lor08,Wu11,Tana11,Fuj12,Susa12,Kof12,Qin15,Qin16,Qin17},
few efforts were found to combine these results with
parameter estimations \cite{Nish12,Ked12}.
In Ref.\ \cite{Nish12},
analysis for frequency-shift measurement based on the Mach-Zehnder interferometer
was presented in the absence of technical noise,
with a particular treatment beyond the AAV limit
and focusing on the quantum shot noise from the photon number fluctuations.
Meanwhile, in Ref.\ \cite{Ked12},
which presents a clear analysis for the imaginary WV measurement
and how the technical noise can be used to enhance the SNR,
numerical results beyond the AAV limit were carried out
to show a reasonable behavior for the noise strength dependence,
while the main theoretical (analytic) analysis was restricted in the AAV limit.

In this work, along the lines of Refs.\ \cite{Ked12,Jor14},
which contain the typical analysis based on the SNR and Fisher information,
we apply the WV treatment beyond the AAV limit
to analyze the WVA measurement for parameter estimation.
Our systematic generalizations are based on
the quantum Bayesian approach (or its variant)
for partial-collapse weak measurement \cite{Kor99,Kor11,BR-1,BR-2},
which allows us to update the system state efficiently
from a specific readout of the meter's variable.
Then, the subsequent postselection of the system state
allows us to construct a {\it joint probability} distribution
by means of the chain-rule in probability theory,
which also enables us to account for the various
technical noises very straightforwardly.
Using the joint probability we are able to carry out,
conveniently and analytically,
the expectation and variance of the postselected
measurement results of the meter's variable,
for the use of the SNR characterization.

The paper is organized as follows.
In Sec.\ II we carry out the generalization in terms of the SNR characterization,
with the results beyond the AAV limit
in close parallel to the compact forms under the AAV treatment.
In Sec.\ III we continue the generalization by computing the Fisher information,
also along the lines of analysis under the AAV limit.
Analogy and difference between the SNR and Fisher information characterizations
will be displayed in connection with the generalized treatment
for finite measurement strength, via examination of the Cram\'er-Rao bound.
We further complete the generalization in Sec.\ IV
by including technical noise and in particular
analyze the remarkable results of imaginary WVA measurements.
In Sec.\ V we summarize the work with brief remarks.

\section{Signal-to-noise ratio characterization}

\subsection{AAV's weak values}

Let us recall briefly some basic aspects of the weak values (WV).
In general, consider two coupled systems
(or the degrees of freedom of a single system),
described by the coupling Hamiltonian $H'=\kappa A B$,
with $\kappa$ the coupling strength, $A$ the {\it system} operator,
and $B$ the operator of the {\it measuring device} (meter).
The weak value $A_w$, defined by a pair of preselected and postselected (PPS)
states of the system, manifests itself as the meter's {\it shift}
in the wave function of the measuring device.

To be more specific, following the AAV's original treatment,
let us take the Stern-Gerlach setup as a concrete model,
which is equivalent in theory to many other systems in lab,
such as the optical beam-deflection measurements in quantum optics
and the circuit-QED architecture in solid-state quantum computation.
In the Stern-Gerlach setup, the electron's trajectory is deflected
when it passing through the inhomogeneous magnetic field.
In this context, we treat
the spin degree of freedom of the electron as the {\it system}
and the spatial one (coordinate or momentum) as the {\it meter},
while their interaction is described by $H'=\kappa P A$,
with $P$ the momentum operator and $A=\sigma_z$ the Pauli operator for the spin.
We assume that the system and the meter are initially prepared as
$|\Psi_T\ra=|i\ra |\Phi\ra$,
where the system state reads
$|i\ra=\alpha |1\ra + \beta |2\ra$,
and the meter's state (the transverse wavefunction of the electron)
is assumed as a Gaussian,
$\Phi(x)=(2\pi \sigma^2)^{-1/4}\exp[-x^2/(4\sigma^2)]$,
with $\sigma$ the width of the wavepacket.
Associated with the coupling Hamiltonian,
the evolution of the entire state is governed
by the unitary operator $U=e^{-id P A}$,
with $d=\int_0^{\tau} dt\, \kappa=\kappa \tau$
($\tau$ is the interacting time).
After the interaction, the entire state becomes entangled and is given by
\bea\label{WF-T}
|\widetilde{\Psi}_T\ra =
\alpha |1\ra |\Phi_1\ra + \beta  |2\ra |\Phi_2\ra \,,
\eea
where the meter's wavefunctions read
\bea\label{Phijx}
\Phi_j(x)=\frac{1}{(2\pi \sigma^2)^{1/4}}
      \exp\left[-\frac{(x-\bar{x}_j)^2}{4\sigma^2} \right] \,.
\eea
with $\bar{x}_{1,2}=\pm d$ the Gaussian centers
shifted by the system states $|1\ra$ and $|2\ra$, respectively.

Under the AAV limit (weak enough measurement),
it can be proved that after postselection with $|f\ra$ for the system state,
the meter's wavefunction is approximately given by \cite{AAV88,AV90}
\bea\label{WV-Px}
\widetilde{\Phi}(x)=\frac{1}{(2\pi \sigma^2)^{1/4}}
\exp\left[-\frac{(x-A_w d)^2}{4\sigma^2} \right]  \,,
\eea
where the multiplicative factor reads
\bea
A_w=\frac{\la f|A|i\ra}{\la f|i\ra} \,.
\eea
This is the well known weak values proposed by AAV.
For the readout of $x$, its average associated with the
new ensemble defined by $|i\ra$ and $|f\ra$ (the PPS states)
is given by $ _f\la x\ra_i=({\rm Re}A_w)\, d$.
Since $A_w$ can be very large
(strongly violating the bounds of the eigenvalues of $A$),
we say then that the signal is {\it amplified},
with the amplification factor defined by
\bea
\eta =|_f\la x\ra_i| \,/\, d  \,.
\eea
In the latter part of this work,
we will use $\eta$ as well to denote the amplification rate
for finite strength of measurements, going beyond the AAV limit.
Finally, we may briefly mention that the AAV's WV is a result of postselection,
with the postselection probabilty
\bea
\gamma = |\la f|i\ra|^2 \,.
\eea
Through the whole work, we will also use $\gamma$ to denote
the postselection probabilty under finite strength measurements.

\subsection{Analysis under the AAV limit}

As mentioned above, with the postselection involved measurement,
the signal of the parameter is amplified as
$\widetilde{d}=({\rm Re}A_w)d$,
which can be much larger than the original $d$.
From the noisy quantum measurement,
both parameters $d$ and $\widetilde{d}$ `hide' in the distribution functions,
say, the original distribution $P(x|d)=|\Phi_1(x)|^2$
and the postselected one $\widetilde{P}(x|\widetilde{d})=|\widetilde{\Phi}(x)|^2$.
Below, we analyze the estimate precision associated with
this weak-value-amplification (WVA) technique.
{\it (i)}
If we use only {\it one output data}
for the estimation of the parameter for each of both cases,
the imprecision is characterized
by the statistical variance of the distribution function.
This is the so-called quantum shot noise.
For instance, for $d$, the imprecision of estimate is characterized by
$\delta^2(\hat{d})=\overline{x^2}-(\overline{x})^2=\sigma^2$,
where $\overline{(\cdots)}$ means the average
defined by the distribution function $P(x|d)$.
Similar characterization applies as well for $\widetilde{d}$,
with $\delta^2(\hat{\widetilde{d}})\equiv \tilde{\sigma}^2=\sigma^2$,
based on the result of \Eq{WV-Px}.
{\it (ii)}
If we use $N$ measurement data, i.e.,
using $\hat{d}=\frac{1}{N}\sum_{j=1}^{N}x_j$ as the estimator,
the estimate precision will be improved as $\delta^2(\hat{d})=\sigma^2/N$.
For the WVA technique, let us suppose $N'$ data survived from the $N$ outputs
by postselection,
and use $\hat{\widetilde{d}}=\frac{1}{N'}\sum_{j=1}^{N'}x_j$ as the estimator.
The estimate precision for $\widetilde{d}$ is characterized by the variance
$\delta^2(\hat{\widetilde{d}})=\sigma^2/N'$.

Therefore, as the most direct characterization for the estimate precision,
we follow Ref.\ \cite{Ked12} to introduce the signal-to-noise ratio (SNR) as
\bea\label{SNR-AAV-1}
R^{(w)}_{S/N} = \frac{\widetilde{d}}{\sigma/\sqrt{N'}}
= \sqrt{\gamma}\eta\, R^{(s)}_{S/N}  \,,
\eea
where $R^{(s)}_{S/N} =\frac{d}{\sigma/\sqrt{N}}$
is the SNR of the standard method,
$\gamma=N'/N$ the success probability of postselection,
and $\eta=\widetilde{d}/d $ the amplification rate of the signal.
Under the AAV limit, we find
$\sqrt{\gamma}\eta = |\la f|A|i \ra| = |A_{fi}|$,
which can approach unity by a proper choice of the PPS states.
For instance, consider $A=\sigma_z$
and choose $|i\ra=|\uparrow\ra_x$ and $|f\ra\simeq |\downarrow\ra_x$
(the eigenstates of $\sigma_x$).
Here we map the two-states system to a spin-1/2 particle described
by the Pauli operators, with the correspondence of
$|1\ra=|\uparrow\ra_z$ and $|2\ra=|\downarrow\ra_z$.
Under such choice for the PPS states,
we see that the SNRs of both schemes are almost the same.

In this context, an interesting comment follows that in the WVA scheme
the postselction keeps only a sub-ensemble of the measurement data,
however it can reach similar estimate precision.
In addition to the technical advantages in practice
\cite{Kwi08,How09a,How09b,How10a,How10b,Lun17},
this feature alone is rather unusual, especially
from the perspective of the Fisher information \cite{Jor14}.
That is, the postselection makes the sub-ensemble of data contain
roughly the same amount of information of the whole ensemble of data.

\subsection{Results beyond the AAV limit}

In order to extend the AAV's treatment to finite strength measurement,
a key element is to update the initial (preselected) state
$\rho_i=|i\ra\la i|$ based on the measurement result $x$.
This can be done by applying the quantum Bayesian approach \cite{Kor99,Kor11,BR-1,BR-2},
or directly using \Eq{WF-T} for the present case.
Conditioned on $x$, we formally denote the update as
$\rho_i \rightarrow \tilde{\rho}(x)$.
Accordingly, the $x$ associated postselection probability
is simply given by $P_x(f)=\la f| \tilde{\rho}(x) |f\ra$.
Based on this, neglecting $x$ (summing all the $x$ which passed the postselection),
we obtain the total postselection probability as \cite{Qin15,Qin16,Qin17}
\bea
\gamma &=&\int dx P_i(x)P_x(f)  \nl
 &=& \rho_{f11}\rho_{i11}
+ \rho_{f22 }\rho_{i22} \nl
&+& 2\, {\rm Re}(\rho^*_{f12}\rho_{i12} )
\, e^{-(\bar{x}_1-\bar{x}_2)^2/8\sigma^2}  \,,
\eea
which is a generalization of the AAV result, $\gamma=|\la f|i\ra |^2$.
In deriving this result, we have used
$P_i(x)=\rho_{i11}|\Phi_1(x)|^2 + \rho_{i22}|\Phi_2(x)|^2$,
and $\rho_f=|f\ra\la f|$ for the postselection state.

Actually, it is desirable to introduce the {\it joint} probability
of getting $x$ and passing the postselection of $|f\ra$
\bea
{\rm Pr}(f;x)= P_i(x)P_x(f) / {\cal N} \,,
\eea
while the normalization factor ${\cal N}$ is just equal to $\gamma$.
Using ${\rm Pr}(f;x)$, ensemble averages of $x$ and $x^2$
can be easily calculated.
First, let us consider the average of $x$, i.e.,
$_{f}\langle x \rangle_{i}=\int dx x {\rm Pr}(f;x)$,
from which the amplification rate of the parameter
is simply given by $\eta=|_{f}\langle x \rangle_{i}|\,/\,d$.
After some algebras, we obtain \cite{Qin15,Qin16,Qin17}
\begin{equation}\label{wv-1}
 \frac{ _{f}\langle x\rangle_{i} }{d}
  = \frac  {{\rm Re}A_w}{1+ {\cal G}\, (|A_{w}|^{2}-1)}
  \equiv \frac{{\rm Re}A_w}{{\cal M}}   \,.
\end{equation}
Here we introduced ${\cal G}=(1-e^{-2g})/2$ and $g=(d/2\sigma)^2$,
which is a suitable parameter to characterize the measurement strength.
We also defined the {\it modification} factor ${\cal M}$,
which clearly reflects the modification effect to the AAV result.
In the limit of {\it extremely weak} measurement,
we have ${\cal G}=g\to 0$.
Then, it seems that we can make the limiting ${\cal M}\to 1$
and return to the AAV result.
However, this is true only for the case
that the AAV weak value $A_w$ is not large enough.
In the regime of the {\it anomalous} AAV effect, i.e., when $A_w\to \infty$,
the ${\cal M}$ factor might be large
and will seriously modify the result,
even in the `extremely' weak measurement regime.

Under the AAV limit, the distribution of the postseleted $x$ is still a Gaussian,
with a shifted center but the same width $\sigma$, as shown by \Eq{WV-Px}.
Now, for finite strength measurement, the distribution ${\rm Pr}(f;x)$
is no longer a Gaussian in general, and may have a different width $\widetilde{\sigma}$.
We therefore calculate $_{f}\langle x^2 \rangle_{i}$,
using the distribution function ${\rm Pr}(f;x)$.
After some algebras, we obtain
\bea\label{var-1}
\widetilde{\sigma}^2 &=& _{f}\langle x^2 \rangle_{i} - (_{f}\langle x \rangle_{i})^2  \nl
&=&  \sigma^2 + d^2 \eta \left(  \frac{|A_w|^2+1}{2{\rm Re}A_w} -\eta \right)  \,.
\eea
For the convenience of later use, we further introduce a {\it width change factor} as
\bea\label{eta-sig}
\eta_{\sigma} = \frac{\widetilde{\sigma}}{\sigma}
=\left[ 1+4g\eta \left(  \frac{|A_w|^2+1}{2{\rm Re}A_w} -\eta \right) \right]^{1/2} \,.
\eea

Precisely in parallel to \Eq{SNR-AAV-1} under the AAV limit,
we introduce the SNR for the weak value measurement with finite strength,
$R^{(w)}_{S/N} = \widetilde{d}/(\widetilde{\sigma}/\sqrt{N'})$,
where $\widetilde{d}=|_{f}\langle x \rangle_{i}|$ is given by \Eq{wv-1}.
We further rescale it as
\bea\label{SNR-2}
R^{(w)}_{S/N}
&=& \left(\frac{d}{\sigma/\sqrt{N}}\right)
\left(\sqrt{\gamma}\, \eta/\eta_{\sigma} \right) \nl
&\equiv&  R^{(s)}_{S/N} \left(\sqrt{\gamma}\, \eta/\eta_{\sigma} \right)   \,,
\eea
where $R^{(s)}_{S/N}$ is the SNR of the standard measurement.
By this way, the SNR comparison of the two measurement schemes
is fully captured by the factor
\bea\label{scale-R}
{\cal R}=\sqrt{\gamma}\, \eta/\eta_{\sigma}  \,.
\eea
In Fig.\ 1, we analyze this factor in detail by numerical plots
of the key variables associated with it.

In Fig.\ 1(a), we show the postselection probability.
In the weak value related application problems,
the most interesting regime is that
by setting the postselection state $|f\ra$ nearly orthogonal to
the initial state $|i\ra$. This will result in {\it anomalous} weak values.
However, with the increase of the measurement strength, the initial state $|i\ra$
will be disturbed more seriously by the measurement backaction.
This makes the disturbed state
no longer nearly orthogonal to the initial state $|i\ra$,
causing thus an increase of the postselection probability
with the measurement strength, as shown in Fig.\ 1(a).

\begin{figure}[h]
\includegraphics[scale=0.47]{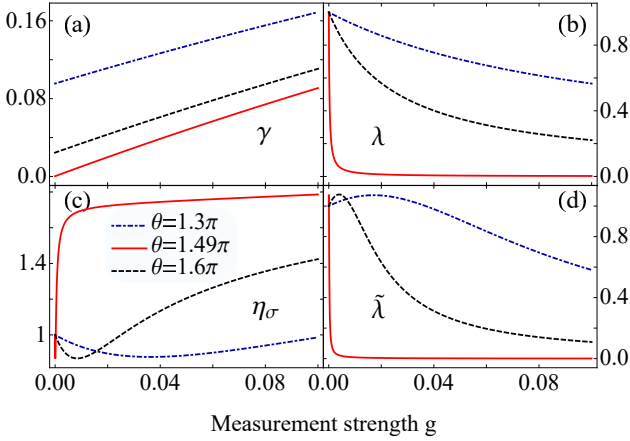}
\caption{
Numerical plots as a function of the measurement strength $g$,
for the a few variables in the SNR scaling factor,
${\cal R}=\sqrt{\gamma}\,\eta/\eta_{\sigma}$, given by \Eq{scale-R}.
In (a) and (b), we show the postselection probability $\gamma$
and the $\lambda$ factor which is defined by $\lambda=\gamma\eta^2/|A_{fi}|^2$,
where $\eta$ is the amplification rate of the signal
and the trivial factor $A_{fi}=\la f|A|i\ra$ is
scaled out for a reason as explained in the main text.
In (c) we plot the width change factor $\eta_{\sigma}=\widetilde{\sigma}/\sigma$,
then in (d) we further include its effect into the scaling factor of the SNR
by introducing $\widetilde{\lambda}=\lambda/\eta^2_{\sigma}$.
All the plots are exemplified by the PPS states $|i\ra=(|1\ra+|2\ra)/\sqrt{2}$
and $|f\ra=\cos\frac{\theta}{2}|1\ra+\sin\frac{\theta}{2}|2\ra$,
with $\theta=1.3\pi$, $1.49\pi$ and $1.6\pi$, respectively.    }
\end{figure}

\begin{figure}[h]
\includegraphics[scale=0.65]{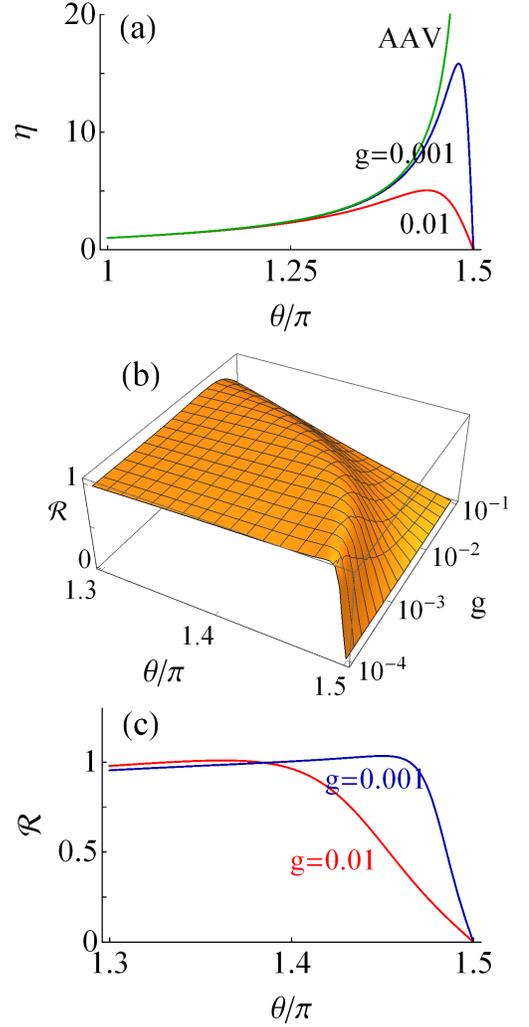}
\caption{
(a)
Deviation of the non-perturbative result, \Eq{wv-1},
from the AAV's prediction when $A_w$ becomes large.
(b)
The overall result of the SNR (scaled by the result of conventional measurement),
while the deviation effect in (a) is manifested here by the `fall-down' area.
(c)
A more evident plot complementary to (b), using two measurement strengths.
Notation of the PPS states $|i\ra$ and $|f\ra$ is the same as in Fig.\ 1.   }
\end{figure}

In the WVA problem, the role of the post-selection is twofold:
it holds the advantage of getting an amplified signal,
while at the same time it suffers the disadvantage
of discarding data, then with a small post-selection probability.
The two competing elements of this type are well described by $\gamma$ and $\eta$.
Under the AAV limit, we simply find $\gamma\eta^2=|\la f|A|i\ra|^2$,
while the trivial factor $A_{fi}=\la f|A|i\ra$
can approximately be unity
by proper design of the initial and post-selection states.
For finite strength measurement, it is not clear
how these two factors compete to each other.
We may thus consider the quantity $\gamma\eta^2$
and in particular to scale out the trivial factor
by introducing $\lambda=\gamma\eta^2/|A_{fi}|^2$.
In the AAV limit, we simply have $\lambda=1$.

In Fig.\ 1(b) we show the behavior of $\lambda$ as a function of
the measurement strength, for a couple of post-selection states.
The common feature is that the factor $\lambda$ decreases from unity
(the limiting value as $g\to 0$) with the increase of the measurement strength,
which indicates a {\it gradual inefficiency} of the WVA technique.
In particular, even for considerably weak measurement strengths,
care is needed for the design of the pre- and post-selection states:
one should not make the AAV weak value $A_w$ too `anomalous',
roughly speaking, which should satisfy the condition $g|A_w|^2<<1$.
The result displayed by the red curve in Fig.\ 1(b)
is an example of this effect,
which indicates a failure of the WVA technique in this case,
despite that the associated AAV weak value $A_w$ is `anomalously' large.
We may remark that this effect cannot be revealed
from the standard AAV result broadly employed in the WVA literature.
Actually, this effect is a consequence of the more rigourous result of \Eq{wv-1},
caused by the denominator of the modification factor ${\cal M}$.
We will return to this point again after a while, with the particular plot of Fig.\ 2.

As we have pointed out, beyond the AAV limit, the distribution
of the postselected data from finite strength of measurement
can considerably deviate from the Gaussian,
and have a different width as shown by \Eq{var-1}.
In Fig.\ 1(c) we plot the width change factor
$\eta_{\sigma}=\widetilde{\sigma}/\sigma$, which is given by \Eq{eta-sig},
as a function of the measurement strength,
while in Fig.\ 1(d) we further include this effect into
the scaling factor $\lambda$ of the SNR
by introducing $\widetilde{\lambda}=\lambda/\eta^2_{\sigma}$.
We find that the modification effect is observable,
especially showing a non-monotonic behavior
with the maximum larger than unity.
However, after accounting for the factor $|A_{fi}|^2$,
as shown in Fig.\ 2,
the signal-to-noise ratio of the WVA measurement
is bounded (approximately) by the result of the standard measurement.

Let us continue the discussion related to the `red curve' in Fig.\ 1(b).
We know that if the PPS states $|i\ra$ and $|f\ra$ are nearly orthogonal,
the {\it anomalous} weak value $A_w$ can violently exceed
the bounds of the eigenvalues of the operator $A$.
For the WVA problem, the standard AAV treatment predicts that
${\rm Re}A_w$ (or ${\rm Im}A_w$) is the amplification rate of the signal.
However, based on a non-perturbative treatment,
the resultant \Eq{wv-1} indicates that the amplification rate
will strongly deviate from the AAV's prediction
when $A_w$ becomes too `anomalously' large.
We clearly show this behavior by the plot of Fig.\ 2(a).

As a consequence of this behavior,
we show in Fig.\ 2(b) the overall result of the SNR
(scaled by the result of the standard measurement),
after accounting for the effects of all the elements of
$\gamma$ (post-selection probability), $\eta$ (signal amplification)
and $\eta_{\sigma}$ (change of distribution width).
The `fall-down' area (from unity to almost zero)
is a direct consequence of the behavior mentioned above for Fig.\ 2(a).
This `fall-down' behavior indicates that, out of our simple expectation,
not a larger $A_w$ will necessarily result in a better effect of amplification.
In practice, one should avoid this area by designing proper post-selection
which, very trickily, depends on the measurement strength.
After avoiding the `dangerous' area,
the results shown in Fig.\ 2(b) and in the complementary Fig.\ 2(c)
display a flat regime with ${\cal R}\simeq 1$,
i.e., reaching nearly the SNR of the standard method.
Actually, in Fig.\ 2(c), we notice that the SNR can even slightly exceed 
the result of the standard measurement,
while the result under the AAV limit 
is strictly bounded by the SNR of the conventional method \cite{Jor14}.  
Again, we emphasize that all the insights gained above
are possible only from the non-perturbative treatment.

\section{Fisher information Characterization}

\subsection{Concept of Fisher information}

In general, for a parameter-$\Omega$ dependent probability distribution of a random 
variable $x$, $P(x|\Omega)$, the Fisher information is defined as \cite{WM09}
\bea\label{FI-1}
{\cal F}(\Omega)=\int dx\, P(x|\Omega)\, [\partial_{\Omega} \ln P(x|\Omega)]^2 \,.
\eea
This is the available {\it information} about the unknown parameter $\Omega$,
or a measure of the sensitivity of $P(x|\Omega)$ to the parameter $\Omega$.
Fisher information is additive. That is, for $N$ independent trials,
the total information is simply given by ${\cal F}_N(\Omega)=N {\cal F}(\Omega)$.
So the most relevant quantity is the Fisher information
extracted by a single probe trial, given by \Eq{FI-1}.
For parameter estimation, the estimator of $\Omega$, denoted as $\hat{\Omega}$,
has the following properties:
{\it (i)} its expectation value satisfies ${\rm E}(\hat{\Omega})=\Omega$;
{\it (ii)} its variance is bounded by the Cram\'er-Rao bound (CRB) as
$\delta^2(\hat{\Omega})\geq 1/{\cal F}(\Omega)$.
This inequality shows that the Fisher information
sets the minimal estimate uncertainty of $\Omega$.

As a little bit extension, if ${\rm E}(\hat{\Omega})\neq\Omega$, i.e.,
the expectation value of the estimator is connected to the {\it original} parameter
via certain functional relation, the CRB inequality reads \cite{WM09}
\bea\label{CRB-1}
[\partial_{\Omega} {\rm E}(\hat{\Omega})]^2 \leq {\cal F}(\Omega)
\, \delta^2(\hat{\Omega}) \,.
\eea
From this result we see that the Fisher information
actually sets the upper bound of the SNR,
by noting that $\delta^2(\hat{\Omega})$ characterizes
the extent of the shot noise of the quantum measurement,
while $\partial_{\Omega} {\rm E}(\hat{\Omega})$ describes
the amplification of the signal (the parameter).
Applying the CRB inequality to the WVA problem,
we may proceed with the following discussions and results.

\subsection{Analysis under the AAV limit}

Let us consider first the standard method.
We identify $\Omega=d$ and the estimator $\hat{\Omega}=\hat{d}$,
which satisfies ${\rm E}(\hat{d})=d$.
Substituting the Gaussian distribution \Eq{Phijx}
into the formula of the Fisher information \Eq{FI-1},
simple calculation yields ${\cal F}=1/\sigma^2$.
Compared with the estimate precision $\delta^2(\hat{d})=\sigma^2$,
we find that the CRB inequality is saturated as an equality,
$\delta^2(\hat{d})=1/{\cal F}$.

Then, let us consider the WVA scheme. We identify $\Omega=d$ and
${\rm E}(\hat{\Omega})={\rm E}(\hat{\widetilde{d}})=\eta d$,
with $\eta=|{\rm Re}A_w|$.
Under the AAV limit, the distribution function of the postselected data
is still a Gaussian, which gives Fisher information as
$\widetilde{{\cal F}}=\eta^2/\sigma^2$.
Again, this result saturates also the CRB inequality,
$\delta^2_w(\hat{d})=1/\widetilde{{\cal F}}$.
Indeed, the Fisher information carried by each postselected data
is enhanced by a factor $\eta^2$, compared to that without postselection.

As done in the SNR characterization,
we further account for the effect of the postselection probability.
For many runs of measurements using $N$ particles,
the total Fisher information of the $N'$ post-selected particles read
\bea
\widetilde{{\cal F}}_{N'}=N' \eta^2/\sigma^2
= \gamma \eta^2 \, {\cal F}_N \,,
\eea
where ${\cal F}_N= N/\sigma^2$ is the total Fisher information
of the $N$ particles without postselection.
Noting that under the AAV limit
$\gamma \eta^2=|\la f|A|i \ra|^2$ (which can approach unity),
we see then, precisely as the SNR discussed in Sec.\ II (B),
that the post-selection makes the sub-ensemble of data ($N'$ particles)
encode roughly the same amount of Fisher information as the whole ensemble
($N$ particles), quite surprisingly, by noting that $N>>N'$.

\subsection{Results beyond the AAV limit}

Going beyond the AAV limit, let us consider the case of
weak measurement with finite strength.
After postselection, the distribution of outputs is largely distorted
from the simple Gaussian, i.e., $P_{f,i}(x)=P_i(x)P_x(f)/{\cal N}$,
with ${\cal N}$ denoting the normalization factor.
Since $P_x(f)$ is $x$ dependent in general,
we know that the new distribution might deviate seriously from $P_i(x)$.
In the above, this postselected distribution has been characterized
by the expectation value and variance of $x$.
Now we further employ the Fisher information to characterize
the effect of postselection. Again, we identify
$\Omega=d$ and ${\rm E}(\hat{\Omega})={\rm E}(\hat{\widetilde{d}})=\eta d$.
Noting that $\eta$ is of $d$ dependence,
we introduce $\widetilde{\eta}=\partial_d(\eta d)=\eta+(\partial_d\eta)d$.
We also denote the variance
$\delta^2(\hat{\widetilde{d}})=\widetilde{\sigma}^2$.
Then, from the CRB inequality we have
\bea\label{CRB-2}
\widetilde{\eta}^2/\eta_{\sigma}^2 \leq \widetilde{{\cal F}}/ {\cal F} \,.
\eea
Here we have used $\eta_{\sigma}=\widetilde{\sigma}/\sigma$
and ${\cal F}=1/\sigma^2$.

\begin{figure}[h]
\includegraphics[scale=0.52]{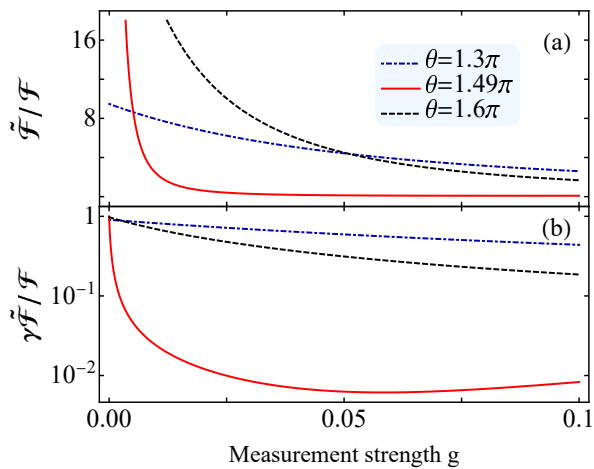}
\caption{
(a) Enhancement of the Fisher information of a single trial,
as a function of the measurement strength.
(b)
Tradeoff result of $\gamma\widetilde{{\cal F}}/ {\cal F}$,
after accounting for the effect of the postselection probability.
The PPS states are assumed the same as in Fig.\ 1.    }
\end{figure}

\begin{figure}[h]
\includegraphics[scale=0.45]{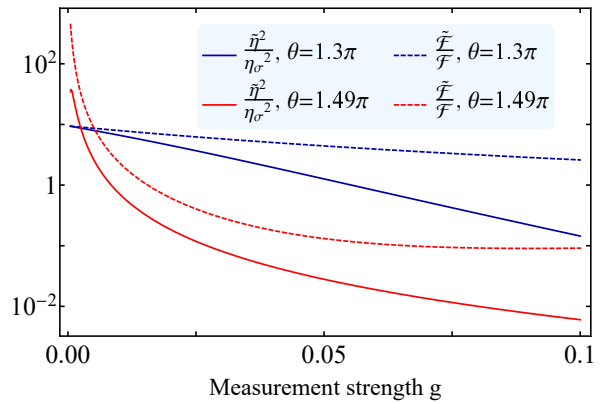}
\caption{
Examination of the CRB inequality,
$\widetilde{\eta}^2/\eta_{\sigma}^2 \leq \widetilde{{\cal F}}/ {\cal F}$,
which is found to be unsaturated
with the increase of the measurement strength.
The PPS states are assumed the same as in Fig.\ 1.   }
\end{figure}

In Fig.\ 3(a), we illustrate the enhancement of the Fisher information
of a single trial, by numerically plotting $\widetilde{{\cal F}}/ {\cal F}$
as a function of the measurement strength.
The reason of the enhancement is also rooted in the new distribution
of the postselected data, which are shifted/amplified by the postselction.
In Ref.\ \cite{Jor14}, this enhancement was highlighted by
that each postselected data contains more Fisher information,
i.e., $\widetilde{{\cal F}}/ {\cal F}>>1$.
However, after accounting for the effect of the postselection probability $\gamma=N'/N$,
we find that the tradeoff result $\gamma\widetilde{{\cal F}}/ {\cal F}$ cannot be larger
than unity, for arbitrary measurement strength and post-selection, as shown in Fig.\ 3(b)

In Fig.\ 4, we further examine the CRB inequality (\ref{CRB-2}) through numerical results.
Under the AAV limit, one can check that \Eq{CRB-2} would reduce as an equality.
However, for the more general case,
we find here that the CRB inequality is becoming unsaturated
with the increase of the measurement strength.
This might be an interesting point in regard to the CRB inequality itself.
The generic reason seems not very obvious,
i.e., from the mathematical derivation of the CRB inequality.
We may leave this issue for possible future investigations.

Roughly speaking, the left-hand-side quantity of \Eq{CRB-2},
$\widetilde{\eta}^2/\eta_{\sigma}^2$,
describes the enhancement of the estimate precision,
in terms of the signal-to-noise ratio.
However, from \Eq{SNR-2}, we find that the precision enhancement factor
is given by $\eta/\eta_{\sigma}$,
with $\eta$ (but not $\widetilde{\eta}$) the signal amplification rate.
If we alternatively use $\eta$ to replace $\widetilde{\eta}$
in the inequality (\ref{CRB-2}), we have found (numerically)
that the inequality (\ref{CRB-2}) cannot be valid in general.
It can be violated for some measurement strength and post-selection.
This indicates that, strictly speaking, the Fisher information characterization
is not precisely equivalent to the characterization
of the signal-to-noise ratio, say, \Eq{SNR-2}.

\section{Effect of Technical Noise}

In the previous analysis of the SNR, the `noise' is actually the quantum uncertainty
of quantum measurement.
In real experiments, there exist other possible technical issues.
In this work we will follow the noise model of Knee and Gauger \cite{Kne14},
which might represent the transverse beam-displacement jitter
in the quantum optical setup and can properly account for the amplifier's noise
in the quadrature measurements of the microwave photons in the circuit QED experiments.
Similar models have also been considered in Refs.\ \cite{How09b,Ste11,Ked12,Jor14}.

\subsection{Measurement in the $x$ basis: $x_0$ noise}

Let us first consider the technical noise $x_0$,
which shifts the meter's wavefunction,
e.g., associated with the initial state $|i\ra$,
from $P_i(x)$ to $P_i(x-x_0)$,
while the noise is assumed to be the typical Gaussian
\bea
{\rm Pr}(x_0)=\frac{1}{\sqrt{2\pi}J} \, e^{-x^2_0/2J^2} \,,
\eea
where $J$ is the width of the noise distribution.
Straightforwardly, the {\it joint probability}
of getting $x$ under the initial state $|i\ra$
and passing the postselection of $|f\ra$,
and as well with the specific noise $x_0$, is given by
\bea
{\rm Pr}(f;x,x_0)=P_i(x-x_0)\, P_{x-x_0}(f)\, {\rm Pr}(x_0)/{\cal N}_f \,,
\eea
where ${\cal N}_f$ is a normalization factor
(given by integrating the variables $x$ and $x_0$).
The two probabilities involved in this result simply read
$P_i(y)=\sum_{j=1,2}\rho_{ijj}P_j(y)$
and $P_{y}(f)=\la f|\widetilde{\rho}(y)|f\ra$,
with a substitution of $y=x-x_0$.
Then, we can carry out any averages under the joint probability through
$ _f\la \bullet\ra_i = \int dx_0 \int dx \, (\bullet)\, {\rm Pr}(f;x,x_0)$.
In particular, the two quantities of our interest are obtained as
\bea
&  _f\la x\ra_i
= \left(\frac{{\rm Re}A_w}{{\cal M}}\right) d  \,,     \nl
&  _f\la x^2\ra_i
= \sigma^2 + J^2 + \left( \frac{\eta d^2}{2} \right)
\left( \frac{1+|A_w|^2}{{\rm Re}A_w} \right)   \,.
\eea
We find that the shift of the signal is not affected by the noise.
However, as expected,
the variance of the amplified signal is added by $J^2$, by noting that
$ _f\la x^2\ra_i -(_f\la x\ra_i)^2 = \widetilde{\sigma}^2 + J^2$
and $\widetilde{\sigma}^2$ is the result shown in \Eq{var-1},
i.e., the variance in the absence of noise.
These two features are the same as in the AAV limit,
despite that the variance $\widetilde{\sigma}^2$
depends on the measurement strength.

\subsection{Measurement in the $p$ basis: $x_0$ noise}

Following Ref.\ \cite{Ked12}, we consider next the more interesting scheme of
the so-called {\it imaginary} WVA technique, which involves the weak measurement
in the $p$ basis, i.e., the eigen-basis of the coupling operator $P$
in the interaction Hamiltonian $H'=\kappa P A$.
The basic idea is as follows. We know that the unitary evolution,
under the action of $U=e^{-id P A}$,
would result in the entangled state of \Eq{WF-T}.
Our previous analysis is based on the measurement in the $x$-basis,
which makes us express the wavefunction of the meter's states as
$\Phi_{1,2}(x)=(2\pi\sigma^2)^{-1/4}\exp[-(x\mp d)^2/4\sigma^2]$.
Now, since we are interested in measurement in the $p$-basis,
after a simple Fourier transformation, the meter's wavefunctions read
$\Phi_{1,2}(p)=(\frac{\pi}{2}\sigma^{-2})^{-1/4}\exp[-\sigma^2 p^2  \mp id\,p]$.
As we will see shortly,
the postselection-associated weak measurement in the $p$-basis
will result in a weak-value-amplification proportional to ${\rm Im}A_w$,
i.e., the imaginary part of the AAV weak value,
which is thus called {\it imaginary} WVA technique.

Further, if the noise is introduced as well by $x_0$,
i.e., a random shift of the meter's wavefunction in the $x$ basis,
the meter's wavefunctions in the $p$ basis can be reexpressed as
\bea
\Phi_{1,2}(p;x_0)=\left(\frac{\pi}{2}\sigma^{-2}\right)^{-1/4}
\exp[-\sigma^2 p^2  \mp id\,p - ix_0\,p] \,.
\eea
Associated with these two wavefunctions, one can check that
the Bayesian approach for state inference does not work.
However, one can update the system state
based on the meter's result of $p$
from the elements of the density matrix $\rho_T=|\Psi_T\ra\la\Psi_T|$,
say, from $\rho_{jk}(p)=\la j|\la p|\rho_T|p\ra|k\ra$ (with $j,k=1$ and 2).
More explicitly, the system state conditioned on the result $p$ is simply given by
\bea
\widetilde{\rho}_{jk}(p)
=\rho_{ijk} \Phi_j(p,x_0)\Phi^*_k(p,x_0)/{\cal N}(p,x_0) \,,
\eea
where the normalization factor reads
${\cal N}(p,x_0)=\sum_{j=1,2}\rho_{ijj}|\Phi_j(p,x_0)|^2$.
We can easily check that the diagonal elements of the system state
remain unchanged under the $p$ basis measurement,
while the off-diagonal elements are updated, for instance, as
\bea
\widetilde{\rho}_{12}(p)=\rho_{i12}\, e^{-i\,2d\,p}  \,.
\eea
Then, the joint probability reads
\bea
{\rm Pr}(f;p,x_0)=P_i(p,x_0)\, P_p(f)\, {\rm Pr}(x_0)/{\cal N}_f \,,
\eea
with $P_i(p,x_0)={\cal N}(p,x_0)$,
$P_p(f)= \la f |\widetilde{\rho}(p)|f\ra$,
and ${\cal N}_f$ a normalization factor.
We find that the both probabilities are free from the noise $x_0$,
knowing thus that any averages of $p$'s functions are free from $x_0$.
Then, an important conclusion is that based on this measurement scheme,
the WVA technique for parameter estimation
can eliminate the negative effect of this type of noise.
Actually, our present result generalizes this claim
from the AAV limit to finite strength of measurement.

\begin{figure}[h]
\includegraphics[scale=0.5]{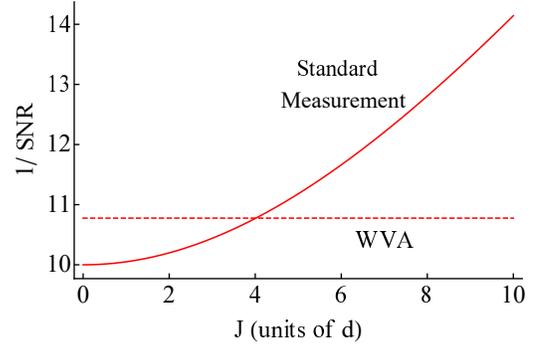}
\caption{
Inverse plot of the SNR,  
for a clearer comparison of $R^{(w)}_{S/N}$ (imaginary WVA measurement)
with $R^{(s)}_{S/N}=d/\sqrt{\sigma^2+J^2}$ (standard measurement), 
while the latter contains the broadening width $J^2$ from the $x_0$ noise,
unlike $R^{(w)}_{S/N}$ which is free from the noise.
The PPS states are chosen as $|i\ra=(|1\ra+|2\ra)/\sqrt{2}$
and $|f\ra=\cos\frac{\theta}{2}e^{i\varphi}|1\ra+\sin\frac{\theta}{2}|2\ra$,
with $\theta=1.49\pi$ and $\varphi=\pi/4$.
The measurement strength in this plot is associated with a choice
of $d=1$ and $\sigma=10$ (in a system of arbitrary units).   }
\end{figure}

Using the joint probability ${\rm Pr}(f;p,x_0)$,
the postselection conditioned averages of $p$ and $p^2$ can be easily obtained
\bea\label{p-meas-1}
& _f\la p\ra_i
= \left( \frac{{\rm Im}A_w}{{\cal M}} \right)
  \left(\frac{d}{2\sigma^2}\right)  e^{-d^2/2\sigma^2}  \,, \nl
& _f\la p^2\ra_i
= \frac{1}{4\sigma^2}  + \left( \frac{|A_w|^2-1}{{\cal M}} \right)
\left(\frac{d^2}{8\sigma^4}\right) e^{-d^2/2\sigma^2}  \,.
\eea
Notice that the first result, say, the new {\it signal}, cannot be regarded as
an amplification of the old signal $d$,
owing to the measurement performed in a different basis.
One can easily check that, for the $p$ basis measurement
and without postselection, the `signal' is zero.
Therefore, in this case, the proper characterization
is the direct use of the SNR in the $p$ basis measurement
\bea\label{Rs/n}
R^{(w)}_{S/N}= \frac{\sqrt{\gamma} \, _f\la p\ra_i}{[\delta^2_w(p)]^{1/2}}  \,,
\eea
where the variance of the postselected result is given by
$\delta^2_w(p) = \,_f\la p^2\ra_i -(_f\la p\ra_i)^2$.
In Fig.\ 5, we make a numerical comparison between
$R^{(w)}_{S/N}$ and $R^{(s)}_{S/N}$,
while the latter is the SNR of the standard measurement,
$R^{(s)}_{S/N}=d/\sqrt{\sigma^2+J^2}$,
which contains the broadening $J^2$ from the $x_0$ noise,
unlike $R^{(w)}_{S/N}$ which is free from the noise.
We find that the imaginary-WVA technique outperforms the standard method
when the noise exceeds certain modest strength,
while both schemes have similar SNR in the absence of noise.

\subsection{Measurement in the $p$ basis: $p_0$ noise}

Let us continue to consider the measurement in the $p$ basis,
but now with a noise caused by a random $p_0$ shift of the $p$ wavepacket.
The noise is assumed as well a Gaussian
\bea
{\rm Pr}(p_0)=\frac{1}{\sqrt{2\pi}J_p} \, e^{-p^2_0/2J_p^2} \,,
\eea
with $J_p$ the width of the noise distribution.
For a specific $p_0$, the meter's wavefunctions are shifted
from $\Phi_{1,2}(p)$ to
$\Phi_{1,2}(p-p_0)=(\frac{\pi}{2}\sigma^{-2})^{-1/4}
\exp[-\sigma^2(p-p_0)^2 \mp i d p]$.
Then, conditioned on the $p$ result of the measurement,
the state can be easily updated as
$\widetilde{\rho}_{12}(p)=\rho_{i12}\, e^{-i\, 2d\, p}$,
while the diagonal elements of the density matrix remain unchanged.

Similarly as above, the joint probability in this case reads
\bea
{\rm Pr}(f;p,p_0)=P_i(p,p_0)\, P_p(f)\, {\rm Pr}(p_0)/{\cal N}_f \,,
\eea
where the initial-state-related probability of getting $p$
is given by $P_i(p,p_0)=\sum_{j=1,2}\rho_{ijj}|\Phi_j(p-p_0)|^2$,
the postselection probability is given by
$P_p(f)= \la f |\widetilde{\rho}(p)|f\ra$,
and the normalization factor ${\cal N}_f$
is given by integrating the variables $p$ and $p_0$.
Accordingly, the postselection conditioned average can be done through
$_f\la \bullet\ra_i = \int dp_0 \int dp \, (\bullet)\, {\rm Pr}(f;p,p_0)$,
which yields
\bea \label{p-meas-2}
& _f\la p\ra_i
= \left( \frac{{\rm Im}A_w}{{\cal M}_k} \right)
\,(2d/\widetilde{\sigma}^2_J) \, e^{-2 d^2/\widetilde{\sigma}^2_J}  \,,   \nl
& _f\la p^2\ra_i
=  1/\widetilde{\sigma}^2_J
 + \left( \frac{|A_w|^2-1} {{\cal M}_k} \right)
 (2d^2/\widetilde{\sigma}^4_J) e^{-2d^2/\widetilde{\sigma}^2_J}  \,.
\eea
Here we introduced the second {\it modification} factor beyond the AAV limit,
${\cal M}_k = 1+ K(|A_w|^2-1)$,
with the measurement strength related factor $K$ given by
$K=(1-e^{-2 d^2/\widetilde{\sigma}^2_J})/2$.
We also introduced an {\it effective width} of uncertainty through
\bea\label{width-Jp}
1/\widetilde{\sigma}^2_J =\frac{1}{4\sigma^2} + J^2_p  \,.
\eea
As above, the first result in \Eq{p-meas-2}
cannot be understood as an amplification of the original signal $d$.
Also, the reasonable characterization in this case
is again using the SNR defined by \Eq{Rs/n}.

In Fig.\ 6, we numerically show the effect of the $p_0$ noise on the SNR.
We may first check that, in the absence of the $p_0$ noise,
for the quantum widths $\sigma=(10, 20, 100)$,
the corresponding SNRs based on the imaginary WVA measurement,
$R^{(w)}_{S/N}=(0.093, 0.046, 0.009)$,
are comparable to that from the standard method,
i.e., $R^{(s)}_{S/N}=(0.1, 0.05, 0.01)$.
Then, after introducing the $p_0$ noise, very strikingly,
we find that one can even use the noise
to increase the estimate precision,
with the increase of $J_p$ until a critical value $J^*_p$.
From Fig.\ 6, taking $\sigma=100$ as an example,
the SNR is enhanced by the noise
by a factor of ${\cal R}\simeq 0.45/0.01=45$.
This is indeed a remarkable result,
in regard to the practical use of the imaginary WVA technique \cite{Ked12}.

\begin{figure}[h]
\includegraphics[scale=0.55]{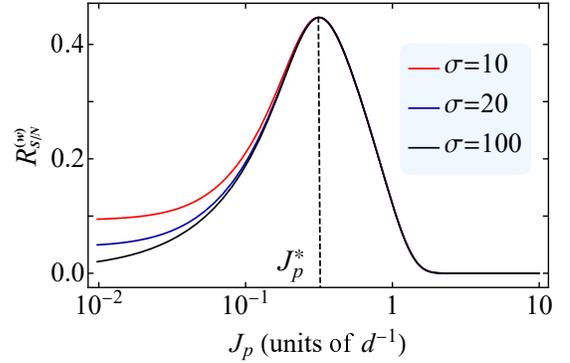}
\caption{
Effect of the $p_0$ noise on the SNR of the imaginary WVA measurement.
In the absence of the $p_0$ noise, the SNR is comparable to
that of the standard method.
Very strikingly, however, after introducing the $p_0$ noise,
the result shows that one can use the noise
to enhance the estimate precision,
with the increase of the noise width $J_p$ until a critical value $J^*_p$.
The same PPS states are chosen as in Fig.\ 5
and the measurement strengths are associated with the choice
of $d=1$ and a few $\sigma$ as shown in the figure
(in a system of arbitrary units).  }
\end{figure}

The critical value of the noise strength, i.e., $J^*_p$ shown in Fig.\ 6,
is determined by the interplay of the a few factors
in the signal-to-noise ratio expression, \Eq{Rs/n}.
First, consider the postselected average $_f\la p\ra_i$.
It has a turnover behavior, qualitatively like that observed in Fig.\ 5.
Indeed, this turnover behavior is dominantly caused by
the two factors,  $2d/\widetilde{\sigma}^2_J$
and $e^{-2 d^2/\widetilde{\sigma}^2_J}$, in \Eq{p-meas-2}.
However, after a careful check, the modification factor ${\cal M}_k$
in $_f\la p\ra_i$ influences also the location of the peak.
Another more prominent influence on the location of the peak
is from the denominator $[\delta^2_w(p)]^{1/2}$ in the signal-to-noise ratio,
which is a monotonically increasing function with $J_p$
and shifts the peak to a smaller value of $J^*_p$.
Finally, the post-selection probability $\gamma$,
which slowly increases with $J_p$,
also has an observable influence on the location of the peak.

We notice that, based on the treatment under the AAV assumptions,
the analytic solution derived in Ref.\ \cite{Ked12}
cannot predict the turnover behavior as shown in Fig.\ 5.
Under the AAV limit, the SNR was found
to be enhanced monotonically by the noise strength $J_p$.
In Ref.\ \cite{Ked12}, the validity condition of the AAV {\it effect}
has been carefully argued, which is needed to ensure
the derived expression of the SNR to be valid.
In order to eliminate the unreasonable prediction at large $J_p$,
numerical results beyond the AAV limit were displayed
in Ref.\ \cite{Ked12} as a necessary correction.
In this context, we may mention that our treatment
and the obtained analytic result \Eq{p-meas-2}
make the SNR be valid (with the turnover behavior) 
without any further modifications.
Finally, we may also mention that the turnover behavior
and the important enhancement of the SNR 
are an overall consequence of 
the type of the interaction Hamiltonian (with the $P$ operator), 
the $p$-basis measurement, the postselection,
and the type of the technical noise introduced in the imaginary WV measurement.
The results are not obvious from a simple intuition,
but nontrivially involve certain complexity of mathematics.

\section{Summary}

We have presented a generalization study for the WVA technique of parameter estimation,
in terms of characterizations of SNR and Fisher information.
Our generalizations were based on the quantum Bayesian approach (or its variant)
for partial-collapse weak measurement with arbitrary strength.
By constructing the joint probability distribution function
associated with postselection and the possible technical noise,
we were able to carry out the various analytical expressions
for the expectation and variance of the postselected
measurement results of the meter's variable.
We thus obtained analytic results of the SNR
and presented systematic analysis
in combination with numerical illustration.

A couple of interesting conclusions may be drawn from our generalized results,
such as that in practice one should avoid too `anomalously' large AAV WV
(in contrast to the naive expectation based on the AAV's treatment),
and can design technical noise strength to achieve optimal SNR
in the imaginary WVA measurement,
as well as that the Fisher information characterization
is not fully equivalent to the SNR characterization
with the increase of measurement strength.
We expect that these conclusions can attract attentions of the WVA community,
from either the experimental perspective or a purely theoretical interest.
We also expect the treatment method
to be applied in the various experimental explorations. \\
\\
\\
\\
\\
{\flushleft\it Acknowledgements.}---
This work was supported by the
National Key Research and Development Program of China
(No.\ 2017YFA0303304) and the NNSF of China (Nos.\ 11675016, 11974011 \& 61905174).

% =====================================================

%\end{CJK*}
\end{document}